\documentclass[conference]{IEEEtran}

\usepackage{cite}
\ifCLASSINFOpdf
   \usepackage[pdftex]{graphicx}
\graphicspath{{./image/}}
     \else
  \usepackage[dvips]{graphicx}
   \DeclareGraphicsExtensions{.eps}
\graphicspath{{./image/}}
\fi
\usepackage{amsmath}
\usepackage{algorithmic}
\usepackage{array}
\ifCLASSOPTIONcompsoc
  \usepackage[caption=false,font=normalsize,labelfont=sf,textfont=sf]{subfig}
\else
  \usepackage[caption=false,font=footnotesize]{subfig}
\fi
\usepackage{stfloats}
\usepackage{url}
\usepackage{amsmath, amssymb}
\usepackage{multirow}
\usepackage{flushend}

\usepackage[left=0.64in,right=0.64in,top=0.8in, bottom=1.1in]{geometry}

\IEEEoverridecommandlockouts
\begin{document}

\title{CacheMamba: Popularity Prediction for Mobile Edge Caching Networks via Selective State Spaces \thanks{This work was partially supported by the Natural Sciences and Engineering Research Council (NSERC) of Canada through the NSERC Discovery Grant RGPIN-2023-05654}}

\author{\IEEEauthorblockN{Ghazaleh Kianfar}
\IEEEauthorblockA{ Concordia Institute of Information\\  Systems Engineering (CIISE)\\
 Concordia University\\
        Montreal, Canada}
\and
\IEEEauthorblockN{Zohreh Hajiakhondi-Meybodi}
\IEEEauthorblockA{ National Research Council Canada \\
        Montreal, Canada}
\and
\IEEEauthorblockN{Arash Mohammadi}
\IEEEauthorblockA{ Concordia Institute of Information\\  Systems Engineering (CIISE)\\
 Concordia University\\
        Montreal, Canada}}

\markboth{Journal of \LaTeX\ Class Files,~Vol.~14, No.~8, August~2015}%
{Shell \MakeLowercase{\textit{et al.}}: Bare Demo of IEEEtran.cls for IEEE Journals}

\maketitle

\begin{abstract}
Mobile Edge Caching (MEC) plays a pivotal role in mitigating latency in data-intensive services by dynamically caching frequently requested content on edge servers. This capability is critical for applications such as Augmented Reality (AR), Virtual Reality (VR), and Autonomous Vehicles (AV), where efficient content caching and accurate popularity prediction are essential for optimizing performance. In this paper, we explore the problem of popularity prediction in MEC by utilizing historical time-series request data of intended files, formulating this problem as a ranking task. To this aim, we propose CacheMamba model by employing Mamba, a state-space model (SSM)-based architecture, to identify the top-$K$ files with the highest likelihood of being requested. We then benchmark the proposed model against a Transformer-based approach, demonstrating its superior performance in terms of cache-hit rate, Mean Average Precision (MAP), Normalized Discounted Cumulative Gain (NDCG), and Floating-Point Operations Per Second (FLOPS), particularly when dealing with longer sequences.
\end{abstract}


%
\IEEEpeerreviewmaketitle

\section{Introduction}
The Sixth Generation (6G) of cellular networks is envisioned to support an exponentially growing user base by integrating advanced technologies, particularly Mobile Edge Caching (MEC) \cite{mahmoud20216g,kianfar2024digital}. MEC strategically places content closer to end-users, thereby significantly reducing the network latency and enhancing its overall performance. In other words, when the requested content is in a nearby cache, the user experiences low latency; otherwise, it faces high latency fetching the request from a remote server. This presents the challenge of developing strategies that adapt to dynamic content popularity while accounting for the limited capacity of edge servers \cite{cheng2023ai}. MEC has several potential applications for advancments of Human Machine Interfaces (HMIs). Generally speaking, MEC is analogous to how recommender systems work, where user preferences and behaviors are analyzed to suggest content that is more likely to be of interest. For example, by caching Augmented Reality (AR) and Virtual Reality (VR)~\cite{pei2023joint} context files (such as 3D models or preprocessed frames) at the edge, MEC allows seamless rendering and interaction, which is critical for real-time immersive experiences. Likewise, in Autonomous Vehicles (AV)~\cite{mao2023ai} operations of vehicle-HMI systems can be minimized by MEC via caching traffic data, map updates, and machine learning models at edge, improving safety and user experience.

\vspace{.025in}
\noindent
\textbf{Literiture Review:} Given the complex nature of population request patterns for various contents, a recent surge of interest has been directed towards implementation of advanced deep learning models for popularity prediction within the comntext of MEC. For instance, Reference~\cite{kang2023content} utilizes a seq2seq Long Short-Term Memory (LSTM) model to predict the number of content requests in a content caching scheme for information-centric networking (ICN). Likewise, in~\cite{li2022spatial} a Temporal Graph Convolutional Network (TGCN) is introduced for predicting content popularity in cellular networks, leveraging spatial-temporal features to enhance prediction accuracy. In addition, \cite{Kianfar2024tdagnn} proposed the Temporal Dual Graph Attention Network (TDA-GNN) for efficient content popularity prediction, addressing the challenge of computational complexity in Graph Neural Network (GNN)-based popularity prediction. To improve generalization and adaptability for predicting content popularity in coded/uncoded content placement scenarios, Reference \cite{meybodi2024multi} developed the Multiple-model Transformer-based Edge Caching (MTEC) framework, which leverages the capabilities of Transformer architecture. 
Despite these advancments, the aformentioned methods face a key limitation, i.e., effectively modeling long sequences for popularity prediction. Models such as LSTM or TGCN, while designed to handle sequential data, can still encounter critical issues with vanishing or exploding gradients when sequences are excessively long. This makes it difficult for these models to learn dependencies over extended time periods, therefore, compromising prediction accuracy. Moreover, the computational demands of models such as TDA-GNN and MTEC increase substantially with longer sequences, leading to longer training times and the necessity for more robust hardware.

Recently, \cite{gu2021efficiently} presented the Structured State Space (S4) model, a novel parameterization of  State-Space Models (SSMs) designed to efficiently handle long-range dependencies in sequence modeling. The S4 model achieves state-of-the-art performance across a variety of benchmarks, including the Long Range Arena tasks, while maintaining computational efficiency and demonstrating competitive results in image and language modeling domains. Building upon the SSMs, Reference~\cite{gu2023mamba} further advances sequence modeling by introducing Mamba, a simplified architecture that integrates selectively parameterized SSMs to replace attention mechanisms. Mamba improves upon S4 by enabling faster inference and linear scalability with sequence length while achieving state-of-the-art performance across various modalities. Since its intorduciton, Mamba has been designed for applicaiton in different domains, such as natural language processing \cite{gu2023mamba}, Human Activity Recognition (HAR) \cite{li2024harmamba}, and sequential recommendation \cite{yang2024uncovering}.

\vspace{.025in}
\noindent
\textbf{Contributions:} 
 To the best of our knowledge, this is the first work to adopt Mamba for popularity prediction in MEC. We investigate the performance of the Mamba model in the context of popularity prediction, with a particular focus on its effectiveness for longer input sequences. To enhance this capability, we propose CacheMamba, a novel framework that leverages the Mamba block to efficiently select the top-$K$ context files with the highest request probabilities. By integrating Mamba’s inherent strengths into our framework, CacheMamba aims to improve popularity prediction by capturing and adapting to evolving request trends while maintaining computational complexity compared to the transformer-based model. Finally, we conduct experiments to evaluate the proposed model's effectiveness, especially in handling longer sequences and dynamic trends in request patterns.

\section{System Model and Problem Formulation}
In this work, we consider a cellular network in which users are served by an edge server. Given that  a distant server stores $N_c$ context files, the goal is to cache top-$K$ requested context files on the edge server to reduce latency in the network. Considering that content placement will be performed during off-peak times (updating time \(t_u\)), the objective is to enhance the cache-hit rate defined as
\begin{equation}
h_{t_u} = \frac{\sum_{i \in \mathcal{S}_{\text{top}}} r^{(W)}_{i,t_u}}{\sum_{j \in \mathcal{S}} r^{(W)}_{j,t_u}}, \label{eq:cache_hit rate}
\end{equation}
where $r^{(W)}_{i,t_u}$ represents the total number of requests for context file $i$ in the time interval with length $W$ between $t_u^{th}$ and $(t_u+1)^{th}$ update times. Additionally, $\mathcal{S}_{top}$ and $\mathcal{S}$ denote the sets of selected top-$K$ and all context files, respectively. 

To select the context files that are likely to maximize cache-hit rate, various time-series forcasting algorithms can be used to predict the top-$K$ popular context files. To this aim, we propose a popularity prediction scheme based on Mamba in the following parts. We initially explain the steps for dataset pre-processing and then proceed with model architecture and cache policy.
\subsection{Dataset Pre-processing}\label{part:dataset}
In this study, we utilize MovieLens-32M dataset, comprising the timestamped ratings of $200,948$ users for $87,585$ movies. The steps to process the dataset are outlined as follows:

\textbf{1) Request Matrix Formation:} We initially sort the requests for context file \(i \in \mathcal{S}\) in ascending temporal order. Assuming the presence of \(T\) timestamps (in seconds), \(\mathbf{R} \in \{0, 1\}^{N_c \times T}\) is defined as an indicator request matrix for each caching node, formulated as follows:

\begin{equation}
\mathbf{R} = 
\begin{bmatrix}
1 & \cdots & 0 \\
0 & \cdots & 1 \\
\vdots & \ddots & \vdots \\
1 & \cdots & 0 \\
\end{bmatrix}_{N_c \times T}
\label{1}
\end{equation}
where each of its elements is denoted by \(r_{l,t}=1\) when context file $l$ is requested at time $t$; otherwise, $r_{l,t} = 0$.
\begin{figure*}[!t]
  \centering
  \subfloat[Mamba architecture.]{\includegraphics[height=8cm]{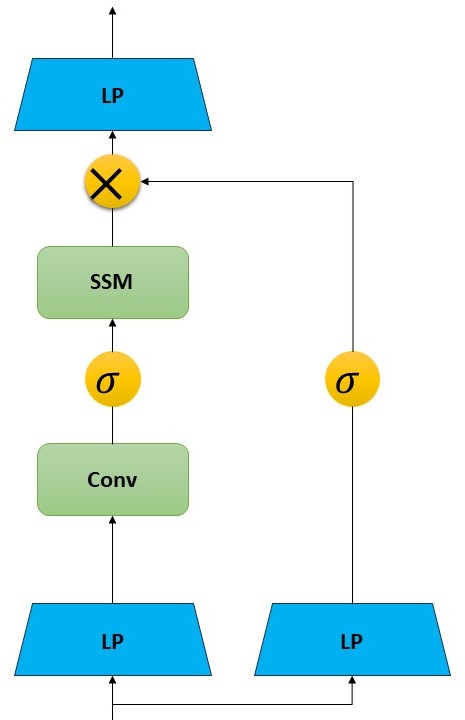}\label{fig:mamba}}
\hspace{1cm}
  \subfloat[CacheMamba model.]{\includegraphics[height=8cm]{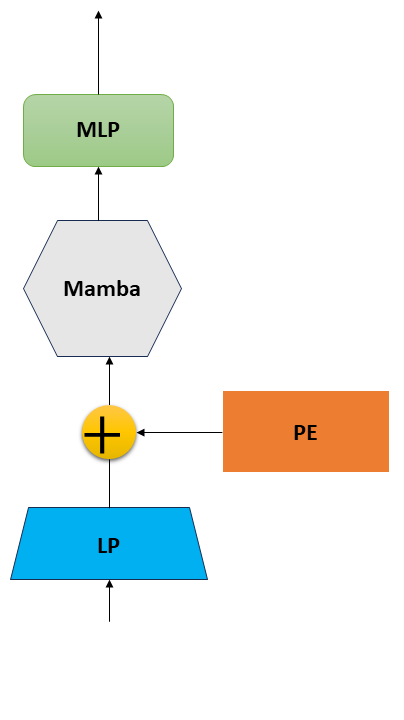}\label{fig:cachemamba}}
  \caption{Illustration of the proposed CacheMamba architecture;(a) Block diagram of Mamba (left), and (b) CacheMamba model (right).}
\end{figure*}

\textbf{2) Time Discretization:} We divide the total duration \(T\) into \(N_w\) discrete intervals, each with a length of \(W\), where \(W\) represents the interval between consecutive updating times. Consequently, a discretized request matrix, denoted by \(\mathbf{R}^{(W)} = \left[\mathbf{r}^{(W)}_{0}, \cdots, \mathbf{r}^{(W)}_{t_u},\cdots, \mathbf{r}^{(W)}_{\lfloor\frac{T}{W}\rfloor}\right] \in {\mathbb{Z}^{+}}^{N \times N_w}\) is constructed, where $\mathbf{r}_{t_u}^{(W)} = \left[r_{1,t_u}^{(W)}, \cdots, r_{N_c,t_u}^{(W)}\right] \in \mathbb{R}^{N_c \times 1}$, and 
\begin{equation}
r^{(W)}_{l,t_u} = \sum_{t=(t_u-1)W+1}^{t_uW} r_{l,t},
\end{equation}
represents the total number of requests for context file $l$ during two consecutive updating times $t_u - 1$ and $t_u$.

\textbf{3) Sample Arrangement}: To predict the Top-$K$ popular context files at updating time \(t_u\), we leverage the historical request patterns during the past update times with length \(L\). Therefore, $M$ input samples in the form $\mathbf{X}_u \in {\mathbb{Z}^{+}}^{N_c\times L}$ are generated by segmenting matrix \(\mathbf{R}^{(W)}\) using an overlapping sliding window of length $L$ before time $t_u$, where \(M = N_w-L\) represents the total number of samples. In addition, we model the top-$K$ selection as a ranking problem and create the output samples as
\begin{equation}
\mathbf{y}_u = \text{softmax}\left(\mathbf{r}_{t_u}^{(W)}\right),
\end{equation}
where \(\sum_{l=1}^{N_c} \mathbf{y}_u[l] = 1,~\forall u\). Finally, the generated dataset is constructed as
\begin{equation}
\mathcal{D} = \Big\{(\mathbf{X}_u, \mathbf{y}_u)\Big\}_{u=1}^M.
\end{equation}
\subsection{Proposed CacheMamba Popularity Prediction}
In this subsection, we present the CacheMamba model, utilizing Mamba to encode time-series data, as follows:
\begin{enumerate}
\item \textbf{Normalization and Compression}: In the first step, the input sample undergoes z-score normalization, ensuring that each feature has a mean of zero and a standard deviation of one. This normalization step is essential as it improves the model's training stability and allows it to learn more effectively from features on a consistent scale. 
Then, to compressing the input data into a more compact representation, the data is linearly projected to a lower-dimensional feature vector using a fully connected layer as
\begin{equation}
    \mathbf{V}_u = \mathbf{W}_p \mathbf{X}_u + \mathbf{b}_p,
\end{equation}
where $\mathbf{W}_p \in \mathbb{R}^{d_{\text{model}} \times N_c}$ and $\mathbf{b}_p \in \mathbb{R}^{d_{\text{model}}}$ are trainable parameters, and $\mathbf{V}_u\in\mathbb{R}^{d_{model}\times L}$ denotes a multi-dimensional sequential data with dimension $d_{model}$ and look-back window length $L$.
\item \textbf{Positional Encoding}: To help the model maintain sequential order information, a sinusoidal non-trainable positional encoding is added to the normalized input data as follows\cite{vaswani2017attention}:
\begin{equation}
\mathbf{Q}_{u}=\mathbf{V}_u+\mathbf{PE}
\end{equation}
where 
\begin{equation}
\mathbf{PE}[{\text{d}, 2\ell}] = \sin\left(\frac{\text{d}}{10000^{\frac{2\ell}{d_{\text{model}}}}}\right),
\end{equation}
\begin{equation}
\mathbf{PE}[{\text{d}, 2\ell+1}] = \cos\left(\frac{\text{d}}{10000^{\frac{2\ell}{d_{\text{model}}}}}\right),
\end{equation}
where $\ell<\frac{L}{2}$, and $d= 0,\cdots, d_{\text{model}}-1$ represents the position in the feature vector. 
\item \textbf{Mamba Encoder}: As depicted in Fig.~\ref{fig:mamba} and Fig.~\ref{fig:cachemamba}, the backbone of our model consists of $N_{layer}$ layers of the Mamba encoder, which utilizes State Space Models (SSMs) for sequence modeling. Unlike previous SSM implementations, Mamba introduces a data-dependent selection mechanism and employs a hardware-optimized parallel algorithm in recurrent mode. These advancements enable efficient and effective sequence modeling, particularly for handling long sequences. After processing the input sequence $\mathbf{Q}_u$ through the Mamba block, we extract the final feature vector from the output sequence, specifically $\mathbf{Q}_u[:, -1]$, and feed it into the subsequent block for further processing.
\item \textbf{Multi-layer Perceptron (MLP)}: In the last layer of the model, an MLP encompassing two fully connected layers are applied: the first has a dimension of $d_{fc}$, and the second projects the feature representations to match the number of context files. After the first fully connected layer, a Rectified Linear Unit (ReLU) activation function is applied, while a softmax function is used in the output layer to produce a probability distribution across the classes.
\item \textbf{Loss Function}: Finally, we train the model using the cross-entropy loss function as
\begin{equation}
    \mathcal{L} = - \sum_{i=1}^{B} \sum_{c=1}^{N_c} \mathbf{y}_{i}[c] \log\Big(\hat{\mathbf{y}}_{i}[c]\Big),
\end{equation}
where $B$ denotes the batch size, and $\hat{\mathbf{y}}_{i}$ represents the predicted probability for context file $c$ in sample $i$. Having trained the model, we sort the context files based on the predicted probabilities for a given sample and cache $K$ ones with the highest probabiities in the sorted list.
\end{enumerate}

\section{Simulation Results}
\begin{table*}[t!]
\centering
\caption{Configuration table for Mamba and Transformer models.}
\begin{tabular}{|l|c|c|c|c|c|}
\hline
\textbf{Model}   & \textbf{Version} & \textbf{$N_{layers}$} & \textbf{$d_{model}$} & \textbf{$d_{state}$} & \textbf{$d_{head}$} \\ \hline
\multirow{3}{*}{CacheMamba} & 0                      & 1                    & 64                  & 16               & -                  \\ \cline{2-6} 
                       & 1                      & 1                    & 64                  & 32               & -                  \\ \cline{2-6} 
                       & 2                      & 2                    & 64                  & 16               & -                  \\ \hline
\multirow{3}{*}{Transformer-based} & 0                & 1                    & 64                  & -                & 8                  \\ \cline{2-6} 
                             & 1                & 1                    & 64                  & -                & 4                  \\ \cline{2-6} 
                             & 2                & 2                    & 64                  & -                & 8                  \\ \hline
\end{tabular}
\label{tab:versions}
\end{table*}
In this section, we develop a simulation environment in Python and prepare the context file popularity data as described in \ref{part:dataset}. To preprocess the dataset, we apply a filtering criterion that selects files requested at least 20 times over a minimum of 200 days. This step reduces the dataset size while ensuring that the selected files contain a sufficient number of requests for training and testing. In addition to evaluating the cache-hit rate as defined in Eq. \eqref{eq:cache_hit rate}, we measure the performance of the proposed CacheMamba model using metrics such as Mean Average Precision (MAP) at $K$, Normalized Discounted Cumulative Gain (NDCG) at $K$, and Floating-Point Operations Per Second (FLOPS), where $K=10$. During training, the learning rate is set to $0.0001$, and the Multi-Layer Perceptron (MLP) in the architecture is configured with 128 neurons. Finally, the proposed CacheMamba model's performance is compared against that of a Transformer-based model, where the Mamba block in the architecture is replaced by a Transformer block. We consider three versions for our proposed model and the transformer-based model with $d_{head}$ heads. Table \ref{tab:versions} illustrates the considered three different versions of each model for performance comparison.
In Table \ref{tab:ap_ndcg}, we compare the performance of CacheMamba and Transformer-based models using two look-back window lengths, both with fewer model parameters. As shown in the table, CacheMamba consistently outperforms the Transformer-based models across all metrics. Specifically, CacheMamba achieves MAP scores of 0.1769 and 0.1877 for $L=200$ and $L=2000$, respectively. These values represent improvements of 0.076 and 0.0181 over the Transformer-based models for the corresponding window lengths, highlighting CacheMamba’s superior ability to predict popular files, particularly for larger window lengths.
Furthermore, CacheMamba achieves higher NDCG values, with improvements of 0.0282 and 0.0856 compared to the Transformer-based models. This demonstrates the effectiveness of CacheMamba in ranking files more accurately. Lastly, in terms of cache-hit rate, the results for both models are relatively similar for $L=200$. However, for $L=2000$, CacheMamba exhibits a significant performance boost, achieving a cache-hit rate approximately 0.01 higher than that of the Transformer-based model.
\begin{table*}[t!]
\centering
\caption{Performance of Mamba and Transformer models across different window sizes and versions.}
\begin{tabular}{|l|c|c|c|c|c|c|}
\hline
\textbf{Model} & \textbf{Window Size} & \textbf{Version} & \textbf{Test MAP@$K$} & \textbf{Test NDCG@$K$} & \textbf{Cache-hit Rate} & \textbf{Model Size (K)} \\ \hline
\multirow{6}{*}{\centering CacheMamba} & \multirow{3}{*}{200}  & 0 & \textbf{0.1769} & \textbf{0.8009} & \textbf{0.7729} & 69.909  \\ \cline{3-7} 
                       &                       & 1 & 0.1639 & 0.7261 & 0.7617 & 76.053  \\ \cline{3-7} 
                       &                       & 2 & 0.1349 & 0.5478 & 0.7588 & 102.549  \\ \cline{2-7} 
                       & \multirow{3}{*}{2000} & 0 & 0.1716 & 0.7261 & 0.8149 & 69.909  \\ \cline{3-7} 
                       &                       & 1 & \textbf{0.1877} & \textbf{0.8027} & \textbf{0.8592} & 76.053  \\ \cline{3-7} 
                       &                       & 2 & 0.1482 & 0.6005 & 0.8238 & 102.549  \\ \hline
\multirow{6}{*}{\centering Transformer-based} & \multirow{3}{*}{200}  & 0 & 0.1654 & \textbf{0.7727} & 0.7412 & 87.253  \\ \cline{3-7} 
                              &                       & 1 & 0.1500 & 0.6316 & \textbf{0.7733} & 87.253  \\ \cline{3-7} 
                              &                       & 2 & \textbf{0.1693} & 0.7544 & 0.7511 & 137.237  \\ \cline{2-7} 
                              & \multirow{3}{*}{2000} & 0 & 0.1634 & \textbf{0.7171} & 0.7786 & 87.253  \\ \cline{3-7} 
                              &                       & 1 & 0.1194 & 0.4916 & 0.5028 & 87.253  \\ \cline{3-7} 
                              &                       & 2 & \textbf{0.1696} & 0.6589 & \textbf{0.8501} & 137.237  \\ \hline
\end{tabular}
\label{tab:ap_ndcg}
\end{table*}
Finally, the achieved cache-hit rate is compared to the number of utilized FLOPS in Fig. \ref{fig:cachehit_flops}. As illustrated, the cache-hit rate improves with increasing window length in both models. However, the computational cost, measured in FLOPS, increases disproportionately for the Transformer-based model, growing 18.34 times when the window length increases from $L=200$ to $L=2000$, compared to a smaller 11.71-fold increase for CacheMamba. Consequently, CacheMamba achieves comparable or superior cache-hit rates while requiring significantly fewer FLOPS than Transformer-based models, especially at larger window lengths. This underscores CacheMamba's scalability and computational efficiency, positioning it as a more effective and resource-efficient alternative to Transformer-based approaches.

\begin{figure}[!t]
  \centering
  \includegraphics[width=\columnwidth]{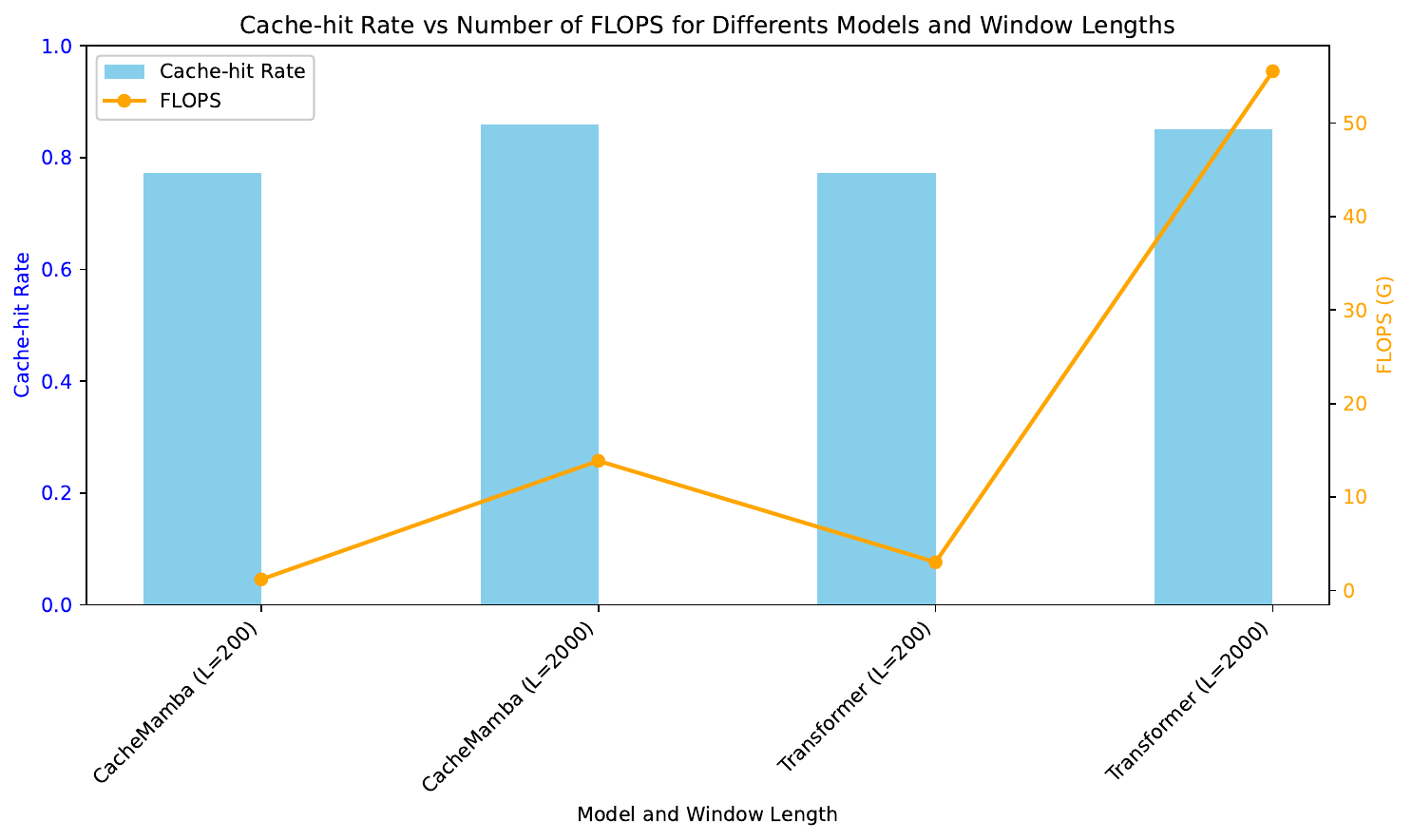}\label{fig:cachehit_flops}
  \caption{Comparing the impact of window length $L$ in CacheMamba and Transformer-based models on cache-hit rate and the number of FLOPS.}
  \label{fig:cachehit_flops}
\end{figure}

\section{Conclusion}
In conclusion, this study addresses the challenge of popularity prediction in Mobile Edge Caching (MEC) by modeling it as a ranking task and utilizing the Mamba architecture, based on state-space models (SSMs), to predict the top-$K$ files most likely to be requested. Our simulations and performance evaluations demonstrate that CacheMamba consistently outperforms Transformer-based models in key metrics, including cache-hit rate, Mean Average Precision (MAP), Normalized Discounted Cumulative Gain (NDCG), and Floating-Point Operations Per Second (FLOPS). CacheMamba excels in prediction accuracy and ranking efficiency, particularly with longer look-back windows, while requiring fewer computational resources. These results highlight the efficiency, scalability, and effectiveness of CacheMamba as a promising solution for popularity prediction in MEC, offering significant improvements in both performance and resource utilization.
 

\bibliographystyle{IEEEtran}
\bibliography{MyBib}

\begin{thebibliography}{10}
\providecommand{\url}[1]{#1}
\csname url@samestyle\endcsname
\providecommand{\newblock}{\relax}
\providecommand{\bibinfo}[2]{#2}
\providecommand{\BIBentrySTDinterwordspacing}{\spaceskip=0pt\relax}
\providecommand{\BIBentryALTinterwordstretchfactor}{4}
\providecommand{\BIBentryALTinterwordspacing}{\spaceskip=\fontdimen2\font plus
\BIBentryALTinterwordstretchfactor\fontdimen3\font minus
  \fontdimen4\font\relax}
\providecommand{\BIBforeignlanguage}[2]{{%
\expandafter\ifx\csname l@#1\endcsname\relax
\typeout{** WARNING: IEEEtran.bst: No hyphenation pattern has been}%
\typeout{** loaded for the language `#1'. Using the pattern for}%
\typeout{** the default language instead.}%
\else
\language=\csname l@#1\endcsname
\fi
#2}}
\providecommand{\BIBdecl}{\relax}
\BIBdecl

\bibitem{mahmoud20216g}
H.~H.~H. Mahmoud, A.~A. Amer, and T.~Ismail, ``6g: A comprehensive survey on
  technologies, applications, challenges, and research problems,''
  \emph{Transactions on Emerging Telecommunications Technologies}, vol.~32,
  no.~4, p. e4233, Feb. 2021.

\bibitem{kianfar2024digital}
G.~Kianfar, S.~J.~S. Dmohammadi, J.~Abouei, A.~Mohammadi, and K.~N.
  Plataniotis, ``Digital aircomp-assisted federated edge learning with adaptive
  quantization,'' in \emph{Proc. 2024 IEEE 4th International Conference on
  Human-Machine Systems (ICHMS)}.\hskip 1em plus 0.5em minus 0.4em\relax IEEE,
  2024, pp. 1--6.

\bibitem{cheng2023ai}
G.~Cheng, C.~Jiang, B.~Yue, R.~Wang, B.~Alzahrani, and Y.~Zhang, ``Ai-driven
  proactive content caching for 6g,'' \emph{IEEE Wireless Communications},
  vol.~30, no.~3, pp. 180--188, June 2023.

\bibitem{pei2023joint}
Y.~Pei, M.~Li, H.~Wu, Q.~Ye, C.~Zhou, S.~Hu, and X.~Shen, ``Joint caching and
  computing resource reservation for edge-assisted location-aware augmented
  reality,'' in \emph{Proc. ICC 2023-IEEE International Conference on
  Communications}.\hskip 1em plus 0.5em minus 0.4em\relax IEEE, 2023, pp.
  2547--2552.

\bibitem{mao2023ai}
B.~Mao, Y.~Liu, J.~Liu, and N.~Kato, ``Ai-assisted edge caching for metaverse
  of connected and automated vehicles: Proposal, challenges, and future
  perspectives,'' \emph{IEEE Vehicular Technology Magazine}, 2023.

\bibitem{kang2023content}
M.~W. Kang and Y.~W. Chung, ``Content caching based on popularity and priority
  of content using seq2seq lstm in icn,'' \emph{IEEE Access}, vol.~11, pp.
  16\,831--16\,842, Feb. 2023.

\bibitem{li2022spatial}
L.~Li, H.~Tian, Y.~Wang, and T.~Zhang, ``Spatial-temporal content popularity
  prediction in cache enabled cellular networks,'' in \emph{Proc. 2022 21st
  International Symposium on Communications and Information Technologies
  (ISCIT)}.\hskip 1em plus 0.5em minus 0.4em\relax IEEE, 2022, pp. 111--116.

\bibitem{Kianfar2024tdagnn}
G.~Kianfar, Z.~Hajiakhondi-Meybodi, and A.~Mohammadi, ``Temporal dual-attention
  graph network for popularity prediction in mec networks,'' in \emph{Proc.
  2024 IEEE 10th World Forum on Internet of Things (WF-IoT)}.\hskip 1em plus
  0.5em minus 0.4em\relax IEEE, 2024.

\bibitem{meybodi2024multi}
Z.~H. Meybodi, A.~Mohammadi, M.~Hou, E.~Rahimian, S.~Heidarian, J.~Abouei, and
  K.~N. Plataniotis, ``Multi-content time-series popularity prediction with
  multiple-model transformers in mec networks,'' \emph{Ad Hoc Networks}, vol.
  157, p. 103436, April 2024.

\bibitem{gu2021efficiently}
A.~Gu, K.~Goel, and C.~R{\'e}, ``Efficiently modeling long sequences with
  structured state spaces,'' \emph{arXiv preprint arXiv:2111.00396}, 2021.

\bibitem{gu2023mamba}
A.~Gu and T.~Dao, ``Mamba: Linear-time sequence modeling with selective state
  spaces,'' \emph{arXiv preprint arXiv:2312.00752}, 2023.

\bibitem{li2024harmamba}
S.~Li, T.~Zhu, F.~Duan, L.~Chen, H.~Ning, C.~Nugent, and Y.~Wan, ``Harmamba:
  Efficient and lightweight wearable sensor human activity recognition based on
  bidirectional mamba,'' \emph{IEEE Internet of Things Journal}, 2024.

\bibitem{yang2024uncovering}
J.~Yang, Y.~Li, J.~Zhao, H.~Wang, M.~Ma, J.~Ma, Z.~Ren, M.~Zhang, X.~Xin,
  Z.~Chen \emph{et~al.}, ``Uncovering selective state space model's
  capabilities in lifelong sequential recommendation,'' \emph{arXiv preprint
  arXiv:2403.16371}, 2024.

\bibitem{vaswani2017attention}
A.~Vaswani, ``Attention is all you need,'' \emph{Advances in Neural Information
  Processing Systems}, 2017.

\end{thebibliography}

\end{document}